# Criticality of the Grad-Shafranov equation: transport barriers and fragile equilibria.

E. R. Solano

Asociación EURATOM-CIEMAT para Fusión, Av. Complutense, 22, E-28040, Madrid, Spain
emilia.solano@ciemat.es

**Abstract**

We review criticality theory as a prelude to consideration of criticality of the Grad-Shafranov equation. Novel criticality conditions of ODEs and PDEs are derived, easily evaluated. The possibility that transport barriers are associated with characteristics of the equilibrium equation is explored. We conjecture that equilibrium criticality permits the appearance of new solution branches: the high confinement branch has high poloidal flux gradient in a diamagnetic region of the plasma. Similarly, criticality may lead to loss of solution, which could be related to MHD instability and/or island formation.

## 1. Introduction

The formation and disappearance of transport barriers in magnetic fusion plasmas is usually considered as a bifurcation in the transport equations. An often considered candidate control parameter for the bifurcation is the local electric field or its shear, which can contribute to turbulence reduction or suppression, as originally proposed by K. C. Shaing [1], experimentally studied for the first time by R. J. Taylor [2] and extensively studied by various authors afterwards [3]. More recent candidates are the local q profile and/or the local magnetic shear, often invoked as controlling parameters for internal transport barrier formation [4].

In this paper, we consider the underlying equilibrium equations, and show that equilibrium criticality may identify suitable starting points for a transport or MHD bifurcation. At a critical point of the Grad-Shafranov equation, a new solution branch of the equilibrium can appear, allowing a step gradient in the poloidal flux ($\nabla\Psi$), which would translate automatically in a steeper gradient of all the flux surface quantities. Equally, a critical point may indicate a fragile solution of the GS equation, susceptible to local or global disruption, or symmetry breaking and subsequent growth of MHD modes or island formation.

First, we briefly introduce concepts from criticality theory in Sections 2. This is followed by a description of the Grad-Shafranov equation, Section 3, and conditions for loss and appearance of solution in Section 4. Finally, we describe in simple physical terms how transport barrier formation and equilibrium criticality may be related, in Section 5.

## 2. Criticality, simple concepts and technical definitions.

What is criticality? An equation for y, such as

$$G(y)=0 \tag{1}$$

is said to be critical or structurally unstable if a small perturbation of the equation (the operator *G*) leads to an essential change in its solution, such as changes in existence and multiplicity of solutions.

The structural stability of polynomial equations is well known [5,6], and it is easily understood graphically. For example, let us consider the cubic equation, $y^3 – (\lambda - \lambda_c) y = 0$, represented in Fig. 1. When $\lambda < \lambda_c$ there is only one solution. For $\lambda = \lambda_c$ there is a critical solution, where the 3 solution branches from the $\lambda > \lambda_c$ region meet. Notice that at the critical equation, the slope of G(y) (considered as a function of y) is horizontal. At the critical point, the cubic equation has G=0, G'(y)=0, G''(y)=0, G'''(y)≠0 (' indicates derivative with respect to y). The cubic is the simplest equation in which new solutions appear. On the other hand, the quadratic equation, illustrated in Fig. 2 provides the simplest example of solution loss, with G=0, G'(y)=0, G''(y)≠0). For polynomial equations, the order of degeneracy of the critical point is given by G'', G''', etc.  Criticality, bifurcation and stability are distinct concepts. Criticality is a qualitative change in the equation properties at $\lambda = \lambda_c$, resulting from perturbation of the equation. Bifurcation is a change from one solution branch to another, only possible when multiple solutions exist, typically controlled by initial or boundary conditions. Bifurcation can happen smoothly at the critical point, or with a "jump" between solutions that are apart. Criticality studies a perturbed equation while stability studies a perturbed solution.

The simple concept of criticality theory, that critically is associated with "zero slope of G(y)", as illustrated in Figs. 1 and 2,, can be generalised to other types of equations, including ODEs and PDEs. A suitably general description of the slope of an operator is its linearization, given by the Gateaux differential in a direction h near a trial function y:

$$G_L(h)\big|_y \equiv \lim_{\varepsilon \to 0} \frac{G(y+\varepsilon h) - G(y)}{\varepsilon} \equiv h\, G_y(y) \tag{2}$$

$G_L(h)\big|_y$ is an operator, function of y, acting on h. $G_y(y) \equiv \partial G/\partial y$ is the Gateaux derivative.

From (2) we find a sufficient condition for criticality at a solution y(x) of equation (1): if in a direction h a non-trivial solution of

$$G_L(h)\big|_y = 0 \tag{3}$$

exists somewhere in the domain of G, then G is critical and y is a critical solution.

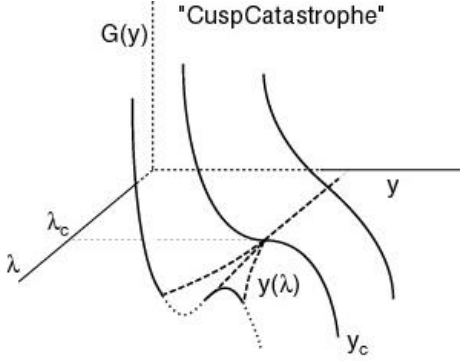
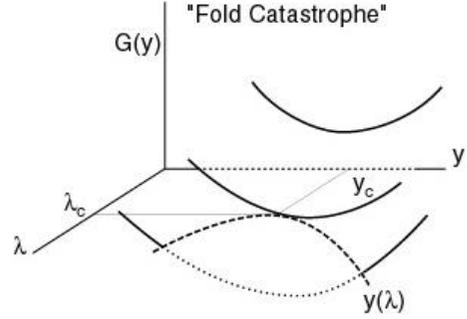

Fig. 1: A cubic equation, displaying a region with 1 solution, smoothly connected to the 3 solution region.

Fig. 2 A quadratic equation, with 0, 1 or 2 solutions.

## 3. Grad Shafranov equation

The Grad-Shafranov (GS) equation is derived from the force-balance equation, $\vec{j} \times \vec{B} = \nabla p$, imposing toroidal symmetry. It can be written as follows:

$$\underbrace{\frac{1}{\mu_0 R}\left(\frac{\partial}{\partial R}\frac{1}{R}\frac{\partial \Psi}{\partial R} + \frac{\partial^2 \Psi}{\partial Z^2}\right)}_{L(\Psi)} + \underbrace{\left(Rp' + \frac{(F^2)'}{2\mu_0 R}\right)}_{J(\Psi)} = 0 \tag{4}$$

We use cylindrical co-ordinates $(R, Z, \zeta)$, with increasing toroidal angle $\zeta$ in the clockwise direction when seen from above. $\vec{B} = F\nabla\zeta + \nabla\zeta \times \nabla\Psi = \vec{B}_\zeta + \vec{B}_{pol}$ is the magnetic field; $\Psi$ is the poloidal magnetic flux per radian, measured from the plasma magnetic axis, outwards; $\vec{j}_\zeta = Rp'R\nabla\zeta + FF'\nabla\zeta/\mu_0$ is the toroidal current density; the prime indicates derivative with respect to $\Psi$; p is the plasma pressure and F the poloidal flux density, both are non-linear functions of $\Psi$ only. The first term of equation (4) defines a linear operator acting on $\Psi$, $L(\Psi)$. The second term defines a generally non-linear operator $J(\Psi)$. $J(\Psi(R,Z))$ is the local (signed) magnitude of the toroidal current density. The poloidal current density is given by $\vec{j}_{pol} = -F' \nabla\Psi \times \nabla\zeta$. If $\vec{j}_{pol}$ adds to the externally driven toroidal field, the plasma is said to be paramagnetic (FF´<0 for a system with positive $\Psi$ and F), and $\vec{j}_{pol} \times \vec{B}_\zeta$ opposes $\vec{j}_\zeta \times \vec{B}_{pol}$. In the opposite case, the plasma is diamagnetic. It is important to point out that the bootstrap current (pressure-gradient driven) is always parallel and therefore paramagnetic, while a pressure gradient also produces a perpendicular current, always diamagnetic. Therefore, for fixed total plasma current, the bootstrap current reduces the capability of the plasma to hold a steep $\nabla p$.

The temporal evolution of the plasma can be described as a succession of equilibrium states, with the plasma transport equations controlling the temporal evolution of p and F as functions of $\Psi$, and of $\Psi(R,Z)$ as a function of time.

## 4. Criticality of Grad-Shafranov equation, loss and appearance of solutions

With the above definitions, the Grad-Shafranov equation and its criticality condition are:

$$L(\Psi) + J(\Psi) = 0 \qquad L(h) + h\, J_\Psi(\Psi) = 0, \qquad (4)$$

If an h exists such that L(h)=0, then $J_\Psi(\Psi) = 0$ indicates criticality at the solution $\Psi$. If, on the other hand, the function h=$\Psi$ satisfies both equations (4), we can derive the following simple sufficient condition for criticality of the equation at a solution $\Psi$:

$$\Psi \frac{\partial J}{\partial \Psi}(\Psi) = J(\Psi) \quad \Leftrightarrow \quad \Psi J_\Psi(\Psi) = J(\Psi) \qquad (5)$$

J and $J_\Psi$ are operators acting on $\Psi(R,Z)$: for each $\Psi(R,Z)$, $J(\Psi)$ is a function of (R,Z), via its explicit dependence on $\Psi$ and R, as is $J_\Psi(\Psi)$. Therefore the criticality conditions (4,5) identify a point, set of points, or flux surface inside the plasma where the equation is critical, and not necessarily a functional form of p´ and FF'. Higher $\Psi$ derivatives of J at the critical points identify the type of criticality. Note that condition (5) is only realisable if J is a non-monotonic function of $\Psi$.

In PDEs, unlike in simple polynomial equations, we must distinguish local and global criticality. Local criticality is given by equation (4) being verified at a point, or in a simply connected set of points in the plasma; it may be sufficient for loss of solution, but can not lead to the appearance of new solutions, since an axi-symmetric equilibrium solution is a global event. Global criticality, present when equation (4) is verified in a closed poloidal loop around the magnetic axis (for instance, but not necessarily, in a whole flux surface), may allow the appearance of new solutions.

In the case of solution loss due to a perturbation of the equation near a fold (local or global), the plasma can no longer be described as ideal, in equilibrium, and toroidally symmetric. The appearance of an MHD mode, or of magnetic islands, or the loss of confinement, local or global, are possible states the plasma may change to. Therefore equation (8) provides a very simple test of MHD: if (8) is verified somewhere with $J_{\Psi\Psi}(\Psi)\neq 0$ the equation has a fold. If the equilibrium evolution, governed by the transport equations, is driven towards the non-solution region, the critical equilibrium is likely to be unstable.

Physically realisable multiple solutions can exist for the plasma equilibrium. They are selected by compatibility with auxiliary constraints: total plasma current, currents in external coils, pressure on axis, etc. Once the plasma is near a globally critical equilibrium, if the transport equations drive the plasma towards the multi-solution region, a small perturbation in the profiles of p or F may lead into a particular solution branch. The difference between the various $\Psi(R,Z)$ equilibria can be understood as a difference in the angle between $\vec{j}$ and $\vec{B}$, since $\vec{j}\times\vec{B}$ supports $\nabla p$, and therefore corresponds to different local diamagnetism and $\nabla\Psi$. Note that a local change in angle can happen in a time scale faster than the field diffusion time scales, which characterise changes in the magnitude of B.

Local criticality at a rational surface may be sufficient for the appearance of an intermediate plasma state in a given flux bundle of the surface, transiently breaking symmetry, and acting as a bridge between two still disconnected neighbouring axi-symmetric solution branches.

## 5. Transport barriers? The equilibrium conjecture.

The simplest example of global criticality of the equilibrium equation leading to new solutions would be given by profiles that somewhere inside the plasma reach a value $\Psi_c$ such that locally they can be expressed as:

$$p'(\Psi)= \alpha_1\Psi, \quad FF'(\Psi)= \gamma_3(\Psi-\Psi_c)^3, \qquad (6)$$

satisfying the criticality condition (5) at the flux surface with $\Psi=\Psi_c$. In that case a perturbation of p' near $\Psi=\Psi_c$ can break the degeneracy of the triple zero and lead to an equation with two new $\Psi(R,Z)$ solution branches. The "high confinement" solution branch would have greater $\nabla\Psi$ in the diamagnetic region, leading to increased $\nabla F$ and therefore greater $\nabla p$, with lower $\nabla\Psi$, $\nabla F$ and $\nabla p$ in the paramagnetic region. The "low confinement" solution would exhibit opposite characteristics. From the critical equilibrium, transport equations and physical constraints dictate the evolution of the system towards one solution or another. For typical Internal Transport Barriers, the paramagnetic region would grow inboard of the diamagnetic one, consistent with the observation of flatter profiles inside the steep gradient region.

We therefore propose the equilibrium conjecture: a transport barrier may appear in a critical equilibrium where FF' or p' have a degenerate zero. After the new solution branches appear, the high confinement branch has, in the "barrier region", steeper gradients of Ψ, and therefore, steeper gradients of pressure, density, temperatures and electric potential. Then the mechanism of turbulence stabilisation due to electric field shear, or Shafranov shift stabilization [7], can be invoked. Note that because the criticality conditions are related to the shape of the current profile, magnetic shear is indirectly connected to the criticality condition.

## 6. Discussion: proof?

Most numerical solvers for the Grad-Shafranov equation are iterative: they oscillate between neighbouring solution branches, and eventually converge to one only, or diverge. It is difficult to study criticality with such tools. We are not aware of analytical solutions of sufficiently non-linear equations. Time-evolving codes usually consider very simplistic J profile functions, usually linear, and usually do not evolve F', but impose its functional dependence on Ψ. Nevertheless, numerical studies of evolution of the equilibrium towards a critical solution and beyond should be possible. Experimentally, equilibrium reconstruction is difficult, although it may be possible, in carefully designed experiments, to establish if FF' has a degenerate zero just before barrier formation. For now, the equilibrium conjecture has no proof, however tantalising its implications may be.

**Acknowledgements:**

H. Wobig pointed out the similarity between the heat transport equation and the Grad-Shafranov equation. G. Spies provided essential mathematical references. This work was funded by the Max-Planck Institut für Plasmaphysick in Garching, the ITER-JCT-Garching, P. H. Edmonds, CIEMAT, JET-EFDA and the Spanish MCyT, via the Ramón y Cajal grant.